# Versatile Millikelvin Hybrid Cooling Platform for Superconductivity Research

Jacob Franklin, Joshua Bedard, and Ilya Sochnikov

*Abstract*—Closed cycle He$^3$-He$^4$ dilution cryostats became the platform of choice in quantum sciences in the era of helium shortage. However, in many experiments, the mechanical vibrations induced by the pulsed cryocoolers present a significant drawback reflected both in electronic and mechanical noises. Here, we present a hybrid dilution cryostat platform; we have automated a commercial closed-cycle system to operate on a cryocooler or on a liquid helium 'battery'. We implemented a scanning SQUID microscope in the hybrid dilution refrigerator. In this work we show the design of the hybrid setup and how its operation eliminates vibration artefacts in magnetic imaging.

## I. Introduction

EXPERIMENTS in Quantum Information Science and Technology mostly operate at extremely low temperatures [1–3]. In solid state devices this is usually a millikelvin range achieved in the He$^3$-He$^4$ dilution refrigerator [4]. Of high interest is the understanding of quantum material properties on the nanoscale [5–7].

The helium shortage and high pricing accelerated the development and adaptation of the so-called 'dry' cryogenic systems, which are widely used for applications such as superconducting qubits [4]. However, using such systems for advanced applications such as microscopy remains limited, in part, due to internal vibrations and noises in the experiments caused by those vibrations.

Several cryogenic scanning probe microscopes have been already realized in closed cycle cryostats. They include scanning gate microscopy [8,9], scanning tunneling microscopy [10], and scanning SQUID microscopy [11–19]. In most of these systems, a rigid scanner or decoupling of vibrations through flexible thermal links or a combination of the two mitigation methods were implemented. Scanning SQUID microscopy is usually done on a scale of hundreds of microns per single image [20–23]. This poses a unique requirement for the fine piezo scanner; it needs to be flexible enough to provide the desired motion range and yet vibrations may be particularly hard to mitigate in such scanners due to their inherit flexibility and the length of the piezo benders. Thus far, vibrations between SQUID sensor and the imaged device or material are at best at 10's of nanometers level. While this level of vibrations is tolerable by micro-SQUIDs [12,24–26], for nano SQUIDs [27–30] significant improvement needs to be realized in vibration mitigation approaches.

## II. Methods

We have implemented a scanning SQUID microscope in a Bluefors LD250 dilution refrigerator, utilizing a custom-built piezoelectric scanner. To mitigate the vibrations which the flexible piezo-benders are susceptible to, we have installed Bluefors' commercially available helium battery (model number BF0916-04), which allows for a hybrid cooling mode in which the typical pulse tube cooler can be shut off for a limited time. While there are many electronics to run associated gauges, pumps, etc., the helium "battery" itself is not electronic. Rather, it is simply a tank of helium bolted to the 4K stage of the dilution refrigerator. This is similar to a liquid helium ballast system [31], but differs in that our device has nothing immersed in the helium but instead provides cooling via contact with the 4K stage of the dilution refrigerator. In this paper we will refer to "charging" the battery as helium flowing into the battery from outside the dilution refrigerator and liquifying, and "discharging" as helium flowing out of the battery and out of the dilution refrigerator and evaporating. The helium battery has a volume of about 1.33L, which can store helium equivalent to about 1000L at NTP (normal temperature and pressure). To facilitate remote operation of the battery and mitigate the consumption of helium, we have constructed a custom-built closed-cycle charging system. A diagram of this system is shown in Figure 1. Helium gas is stored in a large holding tank, which is about 1000 L. This tank is pressurized to about 2 atm (absolute), such that when the battery is charged with 1000L of NTP helium, there is still 1000L of approximately NTP helium left in the tank. The evacuated helium battery, located inside the dilution refrigerator, is charged by opening a path to it and allowing the helium gas to be pulled in and liquified. The battery has a safety limit of 19 PSI, so there is a regulator along this path ensures that the battery is not pressurized past this limit. In addition, a needle valve which is opened slightly ensures that the flow of helium is not so high as to significantly disrupt the temperature of the dilution refrigerator. If contaminants in the helium, such as water vapor or nitrogen gas, get into the battery they will freeze and clog the pipes, rendering the battery unusable until the dilution refrigerator is brought back to room temperature. Thus, a cold trap ensures that any contaminants are frozen out before the gas enters the battery. To discharge the battery, a different path is opened, along which two pumps are used to pump the liquid helium out of the battery and back into the main tank. These pumps are thus always running when the pulse tube is off, but since they are attached to the dilution refrigerator via very long, flexible vacuum bellows, they do not contribute to vibrational noise to our knowledge.

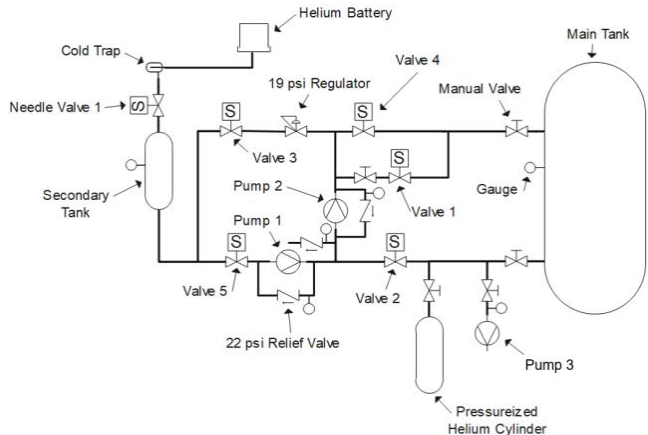

Fig. 1. Diagram of the custom-built closed-cycle automated charging system for the helium battery. The entire system is outside of the dilution refrigerator except for the battery itself. Helium gas is stored in the main tank at around 22 PSI relative pressure. When charging the battery, the helium gas is pulled into the evacuated battery through the top path, through valve 4, the 19 PSI regulator, valve 3, the secondary tank, needle valve 1, and finally the cold trap. Once inside the battery, the helium liquifies. When discharging the battery, the liquid helium is pumped out of the battery, through valve 5, pump 1, pump 2, and finally valve 4 before returning to the main tank. Various gauges and relief valves ensure that the pressure does not increase to unsafe levels anywhere in the system. The elements in the bottom right such as valve 2, the pressurized helium cylinder, and pump 3 are for maintenance uses such as flushing out the helium and refilling the tank if the gas gets contaminated with water vapor or nitrogen.

Photographs of the charging system and the helium battery are shown in Figure 2. The charging system is constructed using standard vacuum bellows and joints but is operated by a custom-built system of solenoid valves that are driven by a USB (universal serial bus) relay connected to a computer. The pumps are also operated through a connection to the computer., which is operated via a remote desktop. All of the components are bolted to or sitting atop a chassis which is comprised of a tall electronics rack with aluminum pieces bolted to it for extra attachment points. The battery is inside the dilution refrigerator in the laboratory, and the charging system is located in a utility room adjacent to the laboratory. Holes in the concrete wall separating the rooms allow for vacuum bellows to transport the helium between the battery and the charging system. The battery takes about 8-10 hours to charge fully and can be performed overnight, as little, if any oversight is needed. The charge lasts for 2-3 hours, during which time the pulse tube may be turned off. This allows for up to 4-6 hours per day of operation without the vibrations of the pulse tube.

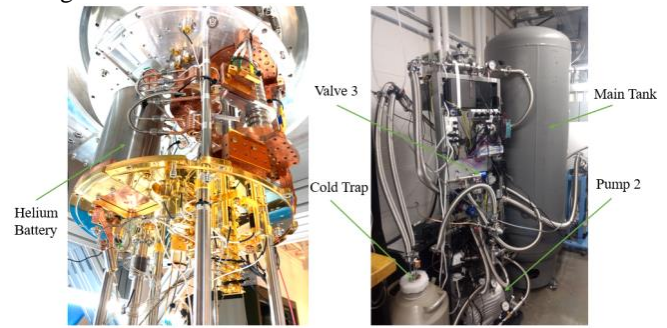

Fig. 2. Photographs of the battery and its charging system. In the left photo, the helium battery can be seen on the 4K stage of the Bluefors dilution refrigerator, which is in the laboratory. The right photo shows the battery charging system, which is located in a utility room directly adjacent to the laboratory. The main components are bolted to a modified electronics rack. A computer on top of the rack is accessed via a remote desktop to control the valves and pumps remotely. Some of the main components which are visible in the view are labelled; please refer to Figure 1 to see how these components contribute to the operation of the battery.

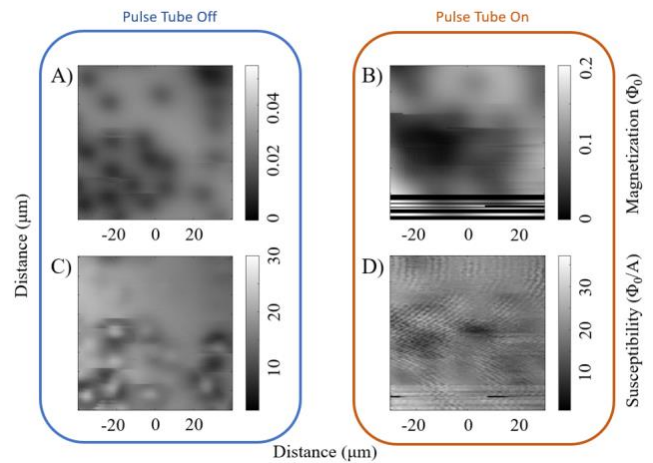

Fig. 3. Scanning SQUID images of superconducting Nb:SrTiO$_3$ (STO) using both the conventional pulse tube cooling and cooling on the helium battery. In panel A), shown is a typical magnetometry image of STO while cooling on the battery, while panel B) shows a typical magnetometry image while on the pulse tube. The black lines are due to a behavior which is typical while using the pulse tube; while the SQUID is scanning in a raster fashion, the vibrations from the pulse tube cause it to experience a "flux jump", which ruins the signal for the remainder of the current line. In panel C), shown is a typical susceptibility image of STO while on the helium battery, and in panel D) shown the same on the pulse tube. The wave-like features in panel D) are the result of periodic noise that we suspect is due to the vibrations of the pulse tube. The spatial resolution of the SQUID is on the order of its pickup loop radius, which is 3 microns. The left images are taken at 100mK about 4 microns above the sample, and the right images are taken at 40mK about 6 microns above the sample. These heights are approximately as close as we can get to the sample for each cooling mode before SQUID-vortex interactions become an issue.

## III. RESULTS

Shown in Figure 3 are scanning SQUID images of superconducting strontium titanate [32] that are typical for the two modes of cooling: the helium battery in panels (A) and (C) and the pulse tube in panels (B) and (D). In both the magnetometry and susceptibility channels the spatial resolution is improved when using the helium battery because the vibrations from the pulse tube makes it so that we must scan with the SQUID a larger distance from the sample. Comparing the magnetometry images, panels (A) and (B), this difference in resolution is clear. In addition, there are very often "flux jumps" in the SQUID when using the pulse tube, which are indicated by the solid black horizontal lines. During these events, which are uncommon when using the helium battery, the vibrations of the SQUID cause a flux quantum to enter or leave the pickup loop of the SQUID, which compromises the integrity of the data for the remainder of the line [33]. In the susceptibility images, panels (C) and (D), we see wave-like features in the image using the pulse tube that indicate a correlated noise. We hypothesize that this noise is due to mechanical vibrations because they are absent when using the helium battery.

## IV. CONCLUSIONS

In conclusion, we have implemented a hybrid dilution refrigerator system for experiments in quantum sciences that require extreme levels of noise rejection. The system consists of a commercial cryostat by Bluefors that is cooled by a cryo-mechanical pulse tube cold head. We installed the helium battery developed by Bluefors in the cryostat, which can provide temporary cooling for 2-3 hours while the pulse tube is idle. We provided details of our custom-built closed-cycle battery operating system, which allows for the charging/discharging of the battery in an automated or remotely controlled manner. We showcase the operation of a scanning SQUID setup with and without the battery to demonstrate that the performance and stability of the microscope are substantially improved with the use of the hybrid mode. To quantify this improvement, spectral analysis of the noise in the scanning SQUID data will be presented in a future work.